\documentclass[showpacs,preprintnumbers,superscriptaddress,amsmath,amssymb,nofootinbib]{revtex4}
\usepackage{graphicx}% Include figure files
\usepackage{bm}% bold math

\usepackage[normalem]{ulem}  % \sout{old text} for strikeout
\usepackage[dvips]{color} % For blue in-text comments and additions

\renewcommand\sout{\bgroup \color{red} \ULdepth=-.5ex \ULset}

\newcommand{\Slash}[1]{\ooalign{\hfil/\hfil\crcr$#1$}}% for Feynman slashs
\newcommand{\Psfig}[2]{\includegraphics[width=#1]{#2}}

\def\mev{\text{ MeV}}
\def\gev{\text{ GeV}}

\def\Kaellen{K\"{a}llen }

\def\trace{\text{tr}}

\def\phph{\phantom{-}}

\begin{document}

\preprint{}

\title{Testing the molecular nature of  $\bm{D_{s 0}^{\ast}(2317)}$  and  $\bm{D_{0}^{\ast} (2400)}$  in semileptonic $\bm{B_s}$ and $\bm{B}$ decays}

\author{Fernando~S.~Navarra} 
\email{navarra@if.usp.br}
\affiliation{Instituto de F\'{\i}sica, Universidade de S\~{a}o Paulo,
  C.P. 66318, 05389-970 S\~{a}o Paulo, SP, Brazil}

\author{Marina~Nielsen}
\email{mnielsen@if.usp.br}
\affiliation{Instituto de F\'{\i}sica, Universidade de S\~{a}o Paulo,
  C.P. 66318, 05389-970 S\~{a}o Paulo, SP, Brazil}

\author{Eulogio~Oset}
\email{oset@ific.uv.es}
\affiliation{ Departamento de F\'{\i}sica Te\'orica and IFIC, Centro
  Mixto Universidad de Valencia-CSIC, Institutos de Investigaci\'on de
  Paterna, Aptdo. 22085, 46071 Valencia, Spain }

\author{Takayasu~Sekihara} 
\email{sekihara@rcnp.osaka-u.ac.jp}
\affiliation{Research Center for Nuclear Physics
  (RCNP), Osaka University, Ibaraki, Osaka, 567-0047, Japan}

\date{\today}% It is always \today, today,

\begin{abstract}

We study the semileptonic $B_s$ and $B$ decays into the  
$D_{s0}^{\ast} (2317)$ and $D_{0}^{\ast} (2400)$ resonances, 
respectively. With the help of a chiral unitarity model in coupled 
channels we compute the ratio of the decay widths of both processes.  
Using current values of the width for the 
$\bar{B}^{0} \to D_{0}^{\ast} (2400)^{+} \,  \bar{\nu}_{l} \, l^{-}$ we make predictions for 
the rate of the 
$\bar{B}_{s}^{0} \to D_{s0}^{\ast} (2317)^{+} \,   \bar{\nu}_{l} \, l^{-}$  decay 
and for the DK invariant mass distribution in  the  
$\bar{B}_{s}^{0} \to D \, K \, \bar{\nu}_{l} \, l^{-}$ decay. 

\end{abstract}

\pacs{%
}% PACS, the Physics and Astronomy
% \keywords{Suggested keywords}%Use showkeys class option if keyword
\maketitle

\section{Introduction}

The recent discovery of many mesons with charm contributed to the
revival of hadron spectroscopy (for recent reviews, see
Ref. \cite{rev}).  Two interesting examples of these mesons are the
$D_{s 0}^{\ast} (2317)$ \cite{d2317} and $D_{0}^{\ast} (2400)$
\cite{d2400} scalar resonances. As it happened to other states, their
measured masses and widths do not match the predictions from
potential-based quark models. This disagreement motivated several
non-conventional (exotic) interpretation of these states.  Among them
the most popular are multiquark configurations in the form of
tetraquarks or meson molecules \cite{rev}.  Since their masses are
located below the $D K$ and $D_s K$ thresholds it is quite natural to
think that they are bound states of $D K$ and $D_s K$ meson pairs. In
the case of the $D_{s 0}^{\ast} (2317)$, additional support to the
molecular interpretation came recently \cite{sasa1} from lattice QCD
simulations.  In all previous lattice studies of the $D_{s 0}^{\ast}
(2317)$, it was treated as a conventional quark-antiquark state and no
states with the correct mass (below the $D K$ threshold) were
found. In Ref. \cite{sasa1}, with the introduction of $D K$ meson
operators, the right mass was obtained.  
In \cite{Liu:2012zya} the scattering length of KD from QCD lattice simulations 
was extrapolated to physical pion masses and then, using the Weinberg compositeness 
condition \cite{weinberg,baru} the dominance of the KD component in the $D_{s0}^*(2317)$ state was concluded.
A reanalysis of the results
of~\cite{sasa1} has been done in \cite{sasa} using the information of
the three energy levels of~\cite{sasa1} and going beyond the effective
range formula. The dominance of the $D K$ component of the
$D_{s0}^{\ast}(2317)$ was firmly established in~\cite{sasa}.  On the
other hand the analysis of the $D_{s 0}^{\ast} (2317) \to D_s^{\ast}
\gamma$ radiative decay suggests that the $D_{s 0}^{\ast} (2317)$ is
most likely an ordinary $\bar{c} s$ state \cite{cola}, although this is disputed in 
\cite{Cleven:2014oka}. It is therefore 
important to carry out further studies to clarify this question. One
aspect to be investigated is the production of $D_{s0}^{\ast}(2317)$
and $D_{0}^{\ast} (2400)$ in $B_s$ and $B$ decays, respectively.

The dominant decay channel of the $B_s$ meson is into the $D_s$ meson plus anything.
Therefore various important properties of the $c \bar{s}$ mesons can be studied in the $B_s$ 
weak decays. In particular, they can shed more light on the controversial $D_{s0}^{\ast}(2317)$
meson, whose nature is still under debate, as discussed above.  
In recent years there has been a significant experimental progress in the  study of the  properties
of the $B_s$ mesons. The Belle Collaboration considerably increased the number of observed
$B_s$ mesons and their decays \cite{belle-bs}. Moreover, $B_s$ mesons are copiously produced at Large Hadron Collider
(LHC) and  precise data on their properties have been taken by the LHCb Collaboration \cite{lhcb-bs}. 
New data are expected in near future \cite{ig-bs}. The study of weak $B_s$ decays is primarily devoted to the
improvement in the determination of the Cabibbo-Kobayashi-Maskawa (CKM) matrix elements, but there are
several interesting topics in hadron physics to be investigated in these processes.

The $D_{s 0}^{\ast} (2317)$ has been measured mostly in B factories
and probably because of its narrow width ($\Gamma < 3.8$ MeV) it has
not been seen in some channels. One of them is the $\bar{B}_s^0$
semileptonic decay, i.e., $\bar{B}_s^0 \rightarrow D_{s0}^{\ast +} \,
\bar{\nu_l} \, l^- $ (Fig. 1).  There are several theoretical
estimates of the branching fraction of this decay channel
\cite{huang,aliev,zhao,li,seg,alb} and they predict numbers which
differ by up to a factor two. In all these calculations, a source of
uncertainty is in the hadronization. In the present work, assuming
that the $D_{s 0}^{\ast} (2317)$ and $D_{0}^{\ast} (2400)$ are
dynamically generated resonances, we try to improve the hadronization
and the treatment of final state meson-meson interactions.  Moreover,
in order to further reduce uncertainties we compute the ratio of decay
widths:
\begin{equation}
R = \frac{\Gamma_{\bar{B}_s^0 \rightarrow D_{s0}^{\ast +}(2317) \, \bar{\nu_l} \, l^-}}
         {\Gamma_{\bar{B}^0 \rightarrow D_{0}^{\ast +}\, \bar{\nu_l} \, l^-}}
\label{ratio}
\end{equation}
We calculate the left side of this equation and then, using the
available experimental information about the process $\bar{B}^0
\rightarrow D_{0}^{\ast} (2400)^{+} \, \bar{\nu_l} \, l^-$ (see
Fig. 2), we can extract $\Gamma_{\bar{B}^0 \rightarrow D_{0}^{\ast}
  (2317)^{+} \, \bar{\nu_l} \, l^-}$.  We consider also the $B^-
\rightarrow D_{0}^{\ast} (2400)^{0} \, \bar{\nu_l} \, l^-$ decay
(Fig. 3).  The formalism is very similar to the one presented in
Ref.~\cite{liang,alberzou} for nonleptonic $B$ decays.

As it is depicted in Fig. 4, after the $W$ emission the remaining $c -
\bar{q}$ pair is allowed to hadronize into a pair of pseudoscalar
mesons (the relative weights of the different pairs of mesons is
known). Once the meson pairs are produced they are allowed to interact
in the way described by chiral unitarity model in coupled channels and
automatically the $D_{s 0}^{\ast} (2317)$ and $D_{0}^{\ast} (2400)$
resonances are produced.
\begin{figure}[ht!]
%\begin{center}
\includegraphics[scale=0.35]{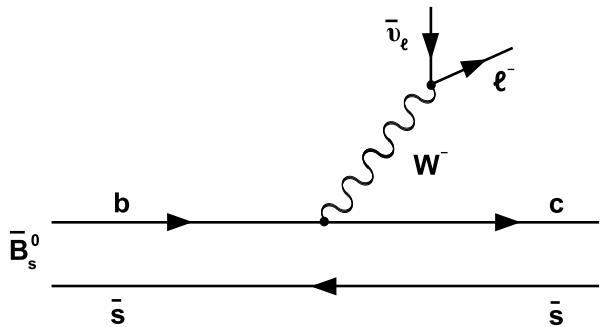} 
\vskip-0.5cm
\caption{Semileptonic decay of $\bar{B}_s^0$ into 
$\bar{\nu}_l l^-$ and a primary $c \bar{s}$ pair.}
%\end{center}
\label{fig1}
\end{figure}
\begin{figure}[ht!]
%\begin{center}
\includegraphics[scale=0.35]{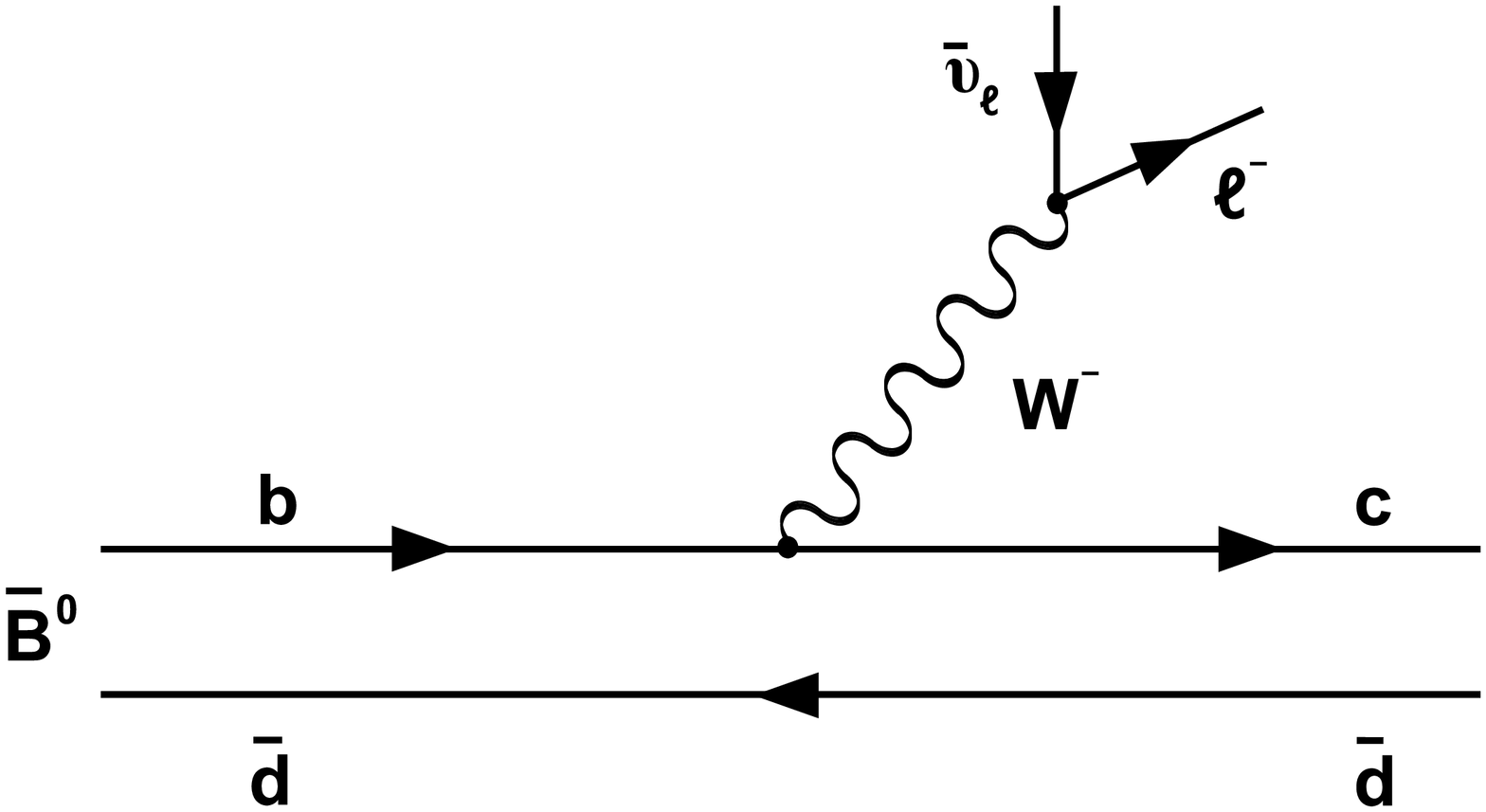} 
\vskip-0.5cm
\caption{Semileptonic decay of $\bar{B}^0$ into $\bar{\nu}_l l^-$ and 
a primary $c \bar{d}$ pair.}
%\end{center}
\label{fig2}
\end{figure}
\begin{figure}[ht!]
%\begin{center}
\includegraphics[scale=0.35]{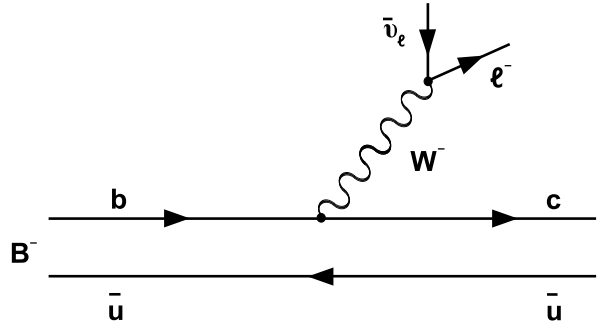} 
\vskip-0.5cm
\caption{Semileptonic decay of $B^-$ into $\bar{\nu}_l l^-$ and 
a primary $c \bar{u}$ pair.}
%\end{center}
\label{fig3}
\end{figure}
%Semileptonic $B$ decays into $p$-wave $D$ resonances $D^{\ast \ast}$,
%$B \to D^{\ast \ast} l \nu$, have been observed in
%Refs.~\cite{Liventsev:2007rb, Aubert:2008ea}.

This paper is organized as follows. In Sec.~\ref{sec:semileptonic}, we
formulate the semileptonic $B$ decay widths into $D$ resonances and
give our model of the hadronization.  Next in
Sec.~\ref{sec:coalescence} we consider $D$ resonance production via
meson coalescence after rescattering, and in
Sec.~\ref{sec:rescattering} we calculate the production of two
pseudoscalars with prompt production plus rescattering through a $D$
resonance.  Then in Sec.~\ref{sec:amplitudes} we formulate meson-meson
scattering amplitudes to generate the $D_{s 0}^{\ast} (2317)$ and
$D_{0}^{\ast} (2400)$ resonances.  In Sec.~\ref{sec:results} we show
our numerical results of the semileptonic $B$ decay widths.
Section~\ref{sec:conclusion} is devoted to drawing the conclusion of
this study.

\section{Semileptonic $\bm{B}$ decays}
\label{sec:semileptonic}

Let us first formulate the semileptonic $B$ decays into $D$ resonances
in the following decay modes:
\begin{equation}
\begin{split}
& \bar{B}_{s}^{0} \to D_{s 0}^{\ast} (2317)^{+} \bar{\nu} _{l} l^{-} , \\
& \bar{B}^{0} \to D_{0}^{\ast} (2400)^{+} \bar{\nu} _{l} l^{-} , \\
& B^{-} \to D_{0}^{\ast} (2400)^{0} \bar{\nu} _{l} l^{-} , 
\end{split}
\end{equation}
where the lepton flavor $l$ can be $e$ and $\mu$.  For this purpose we
express the decay amplitudes and widths in a general form in
Sec.~\ref{sec:decay_width} and give our model of the hadronization in
Sec.~\ref{sec:hadronization}.

\subsection{Semileptonic decay widths}
\label{sec:decay_width}

In general, by using the propagation of the $W$ boson and its
couplings to leptons and quarks, we can express the decay amplitude of
$B \to \bar{\nu} l^{-} \text{hadron(s)}$, $T_{B}$, in the following
manner:
\begin{align}
- i T_{B} = & \overline{u}_{l} i \frac{g_{\rm W}}{\sqrt{2}} \gamma ^{\alpha}
\frac{1 - \gamma _{5}}{2} v_{\nu} \times
\frac{- i g_{\alpha \beta}}{p^{2} - M_{W}^{2}} 
\notag \\ &
\times \overline{u}_{c} i \frac{g_{\rm W} V_{b c}}{\sqrt{2}} \gamma ^{\beta}
\frac{1 - \gamma _{5}}{2} u_{b} \times 
(- i V_{\rm had}) ,
\end{align}
where $u_{l}$, $v_{\nu}$, $u_{c}$, and $u_{b}$ are Dirac spinors
corresponding to the lepton $l^{-}$, neutrino, charm quark, and bottom
quark, respectively, $g_{\rm W}$ is the coupling constant of the weak
interaction, $V_{b c}$ is the Cabibbo-Kobayashi-Maskawa matrix
element, and $M_{W}$ is the $W$ boson mass.  The factor $V_{\rm had}$
consists of the wave function of quarks inside the $B$ meson and the
hadronization contribution in the final state, and it will be
evaluated in the sections below.  In the following we neglect the
squared momentum of the $W$ boson ($p^{2}$) which should be much
smaller than $M_{W}^{2}$ in the $B$ decay process, and therefore the
decay amplitude becomes
\begin{align}
T_{B} = & - i \frac{G_{\rm F} V_{b c}}{\sqrt{2}} 
L^{\alpha} Q_{\alpha} \times V_{\rm had} ,
\label{eq:TB}
\end{align}
where we have introduced the Fermi coupling constant $G_{\rm F} \equiv
g_{\rm W}^{2} / (4 \sqrt{2} M_{W}^{2})$ and defined the lepton and
quark parts of the $W$ boson couplings as.
\begin{equation}
L^{\alpha} \equiv \overline{u}_{l} \gamma ^{\alpha} ( 1 - \gamma _{5} ) v_{\nu} ,
\quad 
%\end{equation}
%\begin{equation}
Q_{\alpha} \equiv \overline{u}_{c} \gamma _{\alpha} ( 1 - \gamma _{5} ) u_{b} ,
\label{eq:LQ}
\end{equation}
respectively.

Now let us calculate the decay widths of the semileptonic $B$ mesons
into $D$ resonances.  In the calculation of the decay widths, we take
the absolute value of the decay amplitude $T_{B}$ and average (sum)
the polarizations of the initial-state quarks (final-state leptons and
quarks).  Therefore, in terms of the amplitude in Eq.~\eqref{eq:TB},
we can obtain the squared decay amplitude as
\begin{equation}
\frac{1}{2} \sum _{\rm pol} | T_{B} |^{2}
= \frac{| G_{\rm F} V_{b c} V_{\rm had} |^{2}}{4}
\sum _{\rm pol} | L^{\alpha} Q_{\alpha} |^{2}
\end{equation}
where the factor $1/2$ comes from the average of the bottom quark
polarization.  Then, by using the conventions of the Dirac spinors and
traces of Dirac matrices summarized in Appendix~\ref{app:1}, we can
calculate the lepton part of the amplitude~\eqref{eq:LQ}, which reads
\begin{align}
\sum _{\rm pol} L^{\alpha} L^{\dagger \beta}
= & \trace \left [ \gamma ^{\alpha} ( 1 - \gamma _{5} )
\frac{\Slash{p}_{\nu} - m_{\nu}}{2 m_{\nu}}
( 1 + \gamma _{5} ) \gamma ^{\beta}
\frac{\Slash{p}_{l} + m_{l}}{2 m_{l}}
 \right ]
\notag \\
= & 2 \frac{p_{\nu}^{\alpha} p_{l}^{\beta} + p_{l}^{\alpha} p_{\nu}^{\beta}
- p_{\nu} \cdot p_{l} g^{\alpha \beta} - i \epsilon ^{\rho \alpha \sigma \beta}
p_{\nu \rho} p_{l \sigma}}{m_{\nu} m_{l}} ,
\end{align}
where $p_{\nu}$ and $p_{l}$ ($m_{\nu}$ and $m_{l}$) are momenta
(masses) of the neutrino and lepton $l^{-}$, respectively.  In a
similar manner, we can calculate the quark part of the amplitude,
which is given by Eq.~\eqref{eq:LQ}
\begin{align}
\sum _{\rm pol} Q_{\alpha} Q_{\beta}^{\dagger}
= & \trace \left [ \gamma _{\alpha} ( 1 - \gamma _{5} )
\frac{\Slash{p}_{b} + m_{b}}{2 m_{b}}
( 1 + \gamma _{5} ) \gamma _{\beta}
\frac{\Slash{p}_{c} + m_{c}}{2 m_{c}}
 \right ]
\notag \\
= & 2 \frac{p_{b \alpha} p_{c \beta} + p_{c \alpha} p_{b \beta}
- p_{b} \cdot p_{c} g_{\alpha \beta} - i \epsilon _{\rho \alpha \sigma \beta}
p_{b}^{\rho} p_{c}^{\sigma}}{m_{b} m_{c}} ,
\end{align}
with the momenta (masses) of the bottom and charm quarks, $p_{b}$ and
$p_{c}$ ($m_{b}$ and $m_{c}$), respectively.  Now we take a heavy
quark limit and assume that the momentum of both the bottom and charm
quarks are zero at the $B$ rest frame.  Then we have $p_{b}^{\mu} =
(m_{b}, \, \bm{0})$ and $p_{c}^{\mu} = (m_{c}, \, \bm{0})$, and the
quark part of the $W$ coupling can be rewritten as
\begin{align}
\sum _{\rm pol} Q_{\alpha} Q_{\beta}^{\dagger}
= 2 \delta _{\alpha \beta} ,
\end{align}
at the $B$ rest frame.  Here we note that this is the delta function
rather than the metric $g_{\alpha \beta}$.  As a consequence, the
square of $L^{\alpha} Q_{\alpha}$ with polarization summation gives
\begin{align}
\sum _{\rm pol} | L^{\alpha} Q_{\alpha} |^{2}
= & 4 \frac{(2 \delta _{\alpha \beta} p_{\nu}^{\alpha} p_{l}^{\beta}
- p_{\nu} \cdot p_{l} \delta _{\alpha \beta} g^{\alpha \beta} )_{B~\text{rest}}}
{m_{\nu} m_{l}}
\notag \\
= & \frac{16 (E_{\nu} E_{l})_{B~\text{rest}}}{m_{\nu} m_{l}} ,
\end{align}
where we have used $\delta _{\alpha \beta} g^{\alpha \beta} = g^{00} +
g^{11} + g^{22} + g^{33} = -2$.  Finally we obtain the squared decay
amplitude:
\begin{equation}
\frac{1}{2} \sum _{\rm pol} | T_{B} |^{2}
= \frac{4 | G_{\rm F} V_{b c} V_{\rm had} |^{2}}{m_{\nu} m_{l}}
(E_{\nu} E_{l})_{B~\text{rest}} .
\label{amp2}
\end{equation}
With the above squared amplitude we can compute the decay width. We
will be interested in two types of decays: three-body decays, such as
$\bar{B}_s^0 \rightarrow D_{s0}^+ \, \bar{\nu_l} \, l^- $, and
four-body decays, such as $\bar{B}_s^0 \rightarrow D^+ \, K^{0} \,
\bar{\nu_l} \, l^- $ and also for the similar $\bar{B}^0$ and $B^-$
initiated processes. As it will be seen, both decay types can be
described by the amplitude $T_{B}$ with different assumptions for
$V_{\rm had}$. The final formulas for 3 and 4-body decays are then
given by:
\begin{equation}
\Gamma _{3} = \frac{4 m_{\nu} m_{l}}{2 m_{B}} \int
d \Phi _{3} \overline{\sum _{\rm pol}} \sum _{\rm pol} | T_{3} |^{2} ,
\label{gamma3}
\end{equation}
\begin{equation}
\Gamma _{4} = \frac{4 m_{\nu} m_{l}}{2 m_{B}} \int
d \Phi _{4} \overline{\sum _{\rm pol}} \sum _{\rm pol} | T_{4} |^{2} ,
\label{gamma4}
\end{equation}
respectively.  In the equations, $m_{B}$, $m_{\nu}$, and $m_{l}$ are
respectively the masses of the $B$ meson, neutrino $\nu$, and lepton
$l$, $T_{3 (4)}$ is the three- (four-) body decay amplitude, and the
summation symbols represent the average of the polarizations in the
initial state and the sum over the polarizations in the final state.
Moreover, the $n$-body phase space $d \Phi _{n}$ has been introduced
as
\begin{equation}
d \Phi _{n} \equiv
\prod _{i=1}^{n}
\left [ \frac{d^{3} p_{i}}{( 2 \pi )^{3}} \frac{1}{2 E_{i}} \right ]
(2 \pi )^{4} \delta ^{4} ( p_{B} - p_{\rm tot} ) ,
\label{eqA:PSn}
\end{equation}
where $E_{i} \equiv \sqrt{\bm{p}_{i}^{2} + m_{i}^{2}}$ is the on-shell
energy of $i$-th particle with its mass $m_{i}$, $p_{B}^{\mu}$ is the
four-momentum of the initial $B$ meson, and $p_{\rm tot}^{\mu}$ is the
sum of the final-state momentum:
\begin{equation}
p_{\rm tot}^{\mu} \equiv \sum _{i=1}^{n} p_{i}^{\mu} ,
\quad
p_{i}^{\mu} = ( E_{i} , \, \bm{p}_{i} ) .
\end{equation}
In order to proceed with the calculation we need a prescription of hadronization, i.e., after the $W$ emission 
in Figs. 1-3 we must specify a way to convert the outgoing quarks into hadrons and compute $V_{\rm had}$. This will be done in the next 
subsection.

\subsection{Hadronization}
\label{sec:hadronization}

The conversion of quarks into hadrons in the final stage of hadron reactions is a long-standing problem which up to now has no definitive 
solution. Since the energies involved are of the order of a few GeV or  less, this is a  non-perturbative process. For particles produced 
in very high energy collisions and with high transverse momentum, we can use fragmentation functions, which are extracted from data phenomenologically 
and then refined with a perturbative QCD treatment. In some case one can develop an approach based on effective Lagrangians \cite{brarec}.  
In the process considered here, in contrast to the high energy case where many particles are produced along with the formed hadron, only one quark-antiquark 
pair is produced and hadronization is mostly a recombination process which binds together the existing quarks. Here we follow \cite{liang}  and  
describe hadronization as depicted in Fig. 4. An extra $q\bar q$ pair with the quantum numbers of the vacuum, $\bar{u} u  + \bar{d} d + \bar{s} s  
+ \bar{c}  c$, is added to the already existing quark pair. The probability of producing the pair is assumed to be given by a number which is the same for 
all light flavors and which will cancel out when taking ratios of decay widths.  
\begin{figure}[ht!]
%\begin{center}
\includegraphics[scale=0.35]{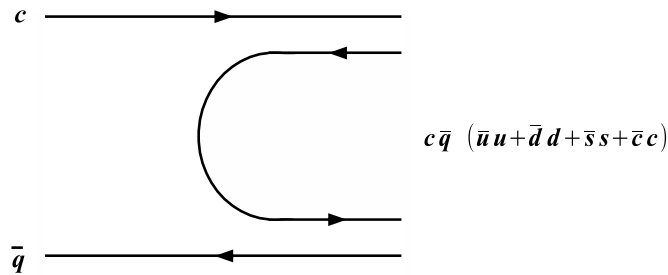} 
%\psfig{file=hqpp_charm_rhic.eps,width=80mm} & \psfig{file=hqpp_bottom_rhic.eps,width=80mm}
%\vskip-0.5cm
\caption{Schematic representation of the hadronization $c\bar q \to c \bar q \, (\bar{u} u + \bar{d} d + \bar{s} s + \bar{c} c)$.}
%\end{center}
\label{fig4}
\end{figure}
We can write this 
$c \bar q \, (\bar{u} u  + \bar{d} d + \bar{s} s  + \bar{c}  c)$  combination 
in terms of pairs of mesons. For this purpose we follow the work of \cite{alberzou} and
define the $q \bar q$ matrix $M$:
\begin{equation}
\label{eq:1}
M=\left(
           \begin{array}{cccc}
             u\bar u & u\bar d  & u\bar s & u\bar c \\
             d\bar u & d\bar d  & d\bar s & d\bar c \\
             s\bar u & s\bar d  & s\bar s & s\bar c \\
             c\bar u & c\bar d  & c\bar s & c\bar c
           \end{array}
         \right)
\end{equation}
which has the property
\begin{equation}
\label{eq:2}
M \cdot M = M \times ( \bar u u + \bar d d + \bar s s + \bar c c ) .
\end{equation}
Now, in terms of mesons the matrix $M$ corresponds to \cite{goz}
\begin{equation}\label{eq:3}
\phi=\left(
  \begin{array}{cccc}
    \frac{1}{\sqrt{2}} \pi^0 + \frac{1}{\sqrt{3}} \eta + 
    \frac{1}{\sqrt{6}} \eta ^{\prime}
    & \pi^+  & K^+ & \bar D^0 \\
    \pi^- &  - \frac{1}{\sqrt{2}} \pi^0 + \frac{1}{\sqrt{3}} \eta + \frac{1}{\sqrt{6}} \eta ^{\prime} & K^0 & D^-       \\
    K^-       & \bar K^0 &  - \frac{1}{\sqrt{3}} \eta + \sqrt{\frac{2}{3}} \eta ^{\prime}    & D_s^- \\
    D^0  & D^+ & D^+_s & \eta_c 
           \end{array}
         \right),
\end{equation}
Hence, in terms of two pseudoscalars we have the correspondence:
\begin{equation}
\label{eq:4}
c\bar s  \,  (\bar{u} u  + \bar{d} d + \bar{s} s  + \bar{c}  c)  \equiv  \left( \phi \cdot \phi \right)_{43} = 
D^0 K^+ + D^+ K^0 + D_s^+ \left ( - \frac{1}{\sqrt{3}} \eta + \sqrt{\frac{2}{3}} \eta ^{\prime} \right ) +  \eta_c D_s^+  
\end{equation}
\begin{equation}
c\bar d  \,  (\bar{u} u  + \bar{d} d + \bar{s} s  + \bar{c}  c)  \equiv  \left( \phi \cdot \phi \right)_{42} = 
D^0 \pi^+ + D^+  \left ( - \frac{1}{\sqrt{2}} \pi^0 + \frac{1}{\sqrt{3}} \eta + \frac{1}{\sqrt{6}} \eta ^{\prime} \right )  +  D_s^+ \bar K^0 + \eta_c D^+   
\label{eq:5}
\end{equation}
\begin{equation}
c\bar u  \,  (\bar{u} u  + \bar{d} d + \bar{s} s  + \bar{c}  c)  \equiv  \left( \phi \cdot \phi \right)_{41} = 
D^0 \left ( \frac{1}{\sqrt{2}} \pi^0 + \frac{1}{\sqrt{3}} \eta + \frac{1}{\sqrt{6}} \eta ^{\prime} \right ) +  D^+ \pi^-  +  D_s^+  K^- + \eta_c D^0   
\label{eq:6}
\end{equation}
for $D_{s 0}^{\ast} (2317)^{+}$, $D_{0}^{\ast} (2400)^{+}$, and
$D_{0}^{\ast} (2400)^{0}$ production, respectively.  Then, for
simplicity we concentrate on the relevant channels for the description
of the $D$ resonances.  In fact, it was pointed out in
Ref.~\cite{Gamermann:2006nm} that the most important channels for the
description of $D_{s 0}^{\ast} (2317)$ ($D_{0}^{\ast} (2400)$) are $D
K$ and $D_{s} \eta$ ($D \pi$ and $D_{s} \bar{K}$).  Therefore, the
weights of the channels to generate the $D$ resonances can be written
in terms of the ket vectors as
\begin{equation}
| ( \phi \phi )_{43} \rangle 
= \sqrt{2} | D K ( 0 , \, 0) \rangle 
- \frac{1}{\sqrt{3}} | D_{s} \eta ( 0 , \, 0) \rangle , 
\end{equation}
\begin{equation}
| ( \phi \phi )_{42} \rangle
= - \sqrt{\frac{3}{2}} | D \pi ( 1/2, \, 1/2) \rangle 
+ | D_{s} \bar{K} ( 1/2, \, 1/2) \rangle ,
\end{equation}
\begin{equation}
| ( \phi \phi )_{41} \rangle
= \sqrt{\frac{3}{2}} | D \pi ( 1/2, \, -1/2) \rangle 
- | D_{s} \bar{K} ( 1/2, \, -1/2) \rangle ,
\label{eq:phiphi_41}
\end{equation}
where we have used two-body states in the isospin basis, which are
specified as $(I, \, I_{3})$ and are summarized in
Appendix~\ref{app:1}.  We note that, due to the isospin symmetry, both
the charged and neutral $D_{0}^{\ast}(2400)$ are produced with the
weight of $| (\phi \phi )_{42} \rangle = - | (\phi \phi )_{41}
\rangle$, which means that the ratio of the decay widths into the
charged and neutral $D_{0}^{\ast}(2400)$ is almost unity.  By using
these weights, we can express $V_{\rm had}$ in terms of two
pseudoscalars.

Once the quark-antiquark pair hadronizes into two mesons they start to
interact and the $D$ resonances can be formed as a result of complex
two-body interactions with coupled channels described by the
Bethe-Salpeter equation.  If the resonance is formed, independent of
how it decays, the process is usually called
``coalescence''~\cite{bracoa} and it is a reaction with three
particles in the final state (see Fig.~5). If we look for a specific
two meson final channel we can have it by ``prompt'' or direct
production (first diagram of Fig.~6), and by rescattering, generating
the resonance (second diagram of Fig.~6). This process is usually called 
``rescattering'' and it is a reaction with four particles in the final
state. Coalescence and rescattering will be discussed in the next
sections.

\begin{figure}[t!]
%\begin{center}
\includegraphics[scale=0.35]{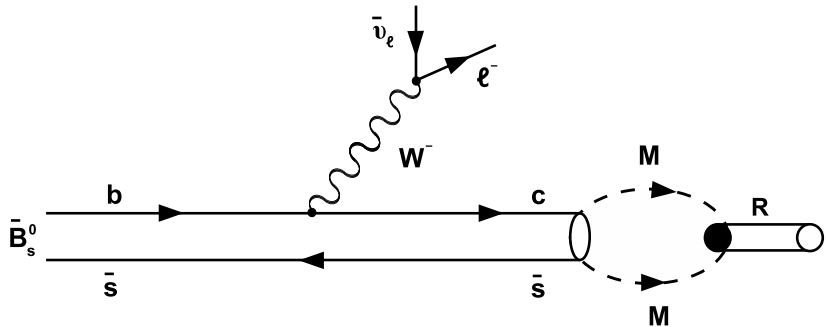}
\vskip-2.5cm
\caption{Diagrammatic representation of $ D_{s0}^{\ast +} (2317)$ production via meson coalescence after rescattering.}
%\end{center}
\label{fig5}
\end{figure}
\begin{figure}[t!]
%\begin{center}
\includegraphics[scale=0.25]{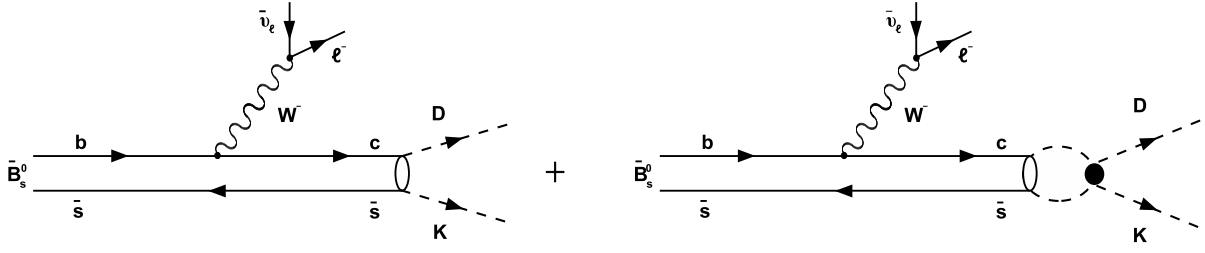}
%\psfig{file=hqpp_charm_rhic.eps,width=80mm} & \psfig{file=hqpp_bottom_rhic.eps,width=80mm}
%\vskip0.5cm
\caption{Diagrammatic representation of $D K$ production: directly (on the left) and via rescattering (on the right) in $\bar B^0_s$ decays.}
%\end{center}
\label{fig6}
\end{figure}

\section{Coalescence}
\label{sec:coalescence}

In this section we consider $D$ resonance production via meson coalescence after rescattering
(see Fig.~5).
%and the other is the production of two pseudoscalars
%which gives the  peak of the $D$ resonance in the invariant mass distribution (see Fig.~6).  
This process has a three-body final state with a lepton, its neutrino and the resonance R. The 
hadronization factor, $V_{\rm had}$,  can be obtained as
\begin{equation}
  V_{\text{had}} ( D_{s 0}^{\ast}(2317) ) = C
  \left ( \sqrt{2} G_{D K} g_{D K} 
  - \frac{1}{\sqrt{3}} G_{D_{s} \eta} g_{D_{s} \eta} \right ) ,
\label{vhads}
\end{equation}
\begin{equation}
  V_{\text{had}} ( D_{0}^{\ast}(2400)^{+} ) = 
  - V_{\text{had}} ( D_{0}^{\ast}(2400)^{0} ) = C
  \left ( - \sqrt{\frac{3}{2}} G_{D \pi} g_{D \pi} 
  + G_{D_{s} \bar{K}} g_{D_{s} \bar{K}} \right ) ,
\label{vhado}
\end{equation}
Here $g_{i}$ is the coupling constant of the $D$ resonance to the
$i$-th two meson channel and $G_{i}$ is the loop function of two meson
propagators (see Sec.~\ref{sec:amplitudes})
\begin{equation}
G_{i} ( s ) \equiv 
i \int \frac{d ^{4} q}{(2 \pi)^{4}} 
\frac{1}{q^{2} - m_{i}^{2} + i \epsilon } 
\frac{1}{(P - q)^{2} - m_{i}^{\prime 2} + i \epsilon} ,
\label{eq:Gloop}
\end{equation}
where $P^{\mu}$ is the total four-momentum of the two-meson system,
and thus $P^{2} = s$ with $s$ being the invariant mass squared of the
two-meson system, and $m_{i}$ and $m_{i}^{\prime}$ are the masses of
the two mesons in channel $i$. An important point to note is that the
prefactor $C$ is the same in all decay modes and contains dynamical
factors common to all reactions only because we are assuming that in
the hadronization the SU(3) flavor symmetry is reasonable, i.e., the
quark pairs $c \bar{s}$ (Fig.~1) and $c \bar{d}$ (Fig.~2) hadronize in
the same way.  We further assume that $C$ is a constant and therefore
is canceled when we take the ratio of decay widths as in
Eq.~\eqref{ratio}.

Now we can evaluate the decay widths by using the formula of
Eq.~\eqref{gamma3}.  Inserting Eq.~\eqref{vhads} [or
Eq.~\eqref{vhado}] into Eq.~\eqref{amp2} and the latter into
Eq.~\eqref{gamma3} we can write the decay width of the $D$ resonance
production via meson coalescence (Fig.~5) as
\begin{equation}
\Gamma _{\rm coal} = \frac{m_{\nu} m_{l}}{128 \pi ^{5} m_{B}^{2}}
\int d M_{\rm inv}^{( \nu l)} 
p_{D}^{\rm cm} \tilde{p}_{\nu} \int d \Omega _{D} \int d \tilde{\Omega}_{\nu}
\frac{4 | G_{\rm F} V_{b c} V_{\rm had} (D^{\ast}) |^{2}}{m_{\nu} m_{l}}
(E_{\nu} E_{l})_{B~\text{rest}}  
\end{equation}
where $p_{D}^{\rm cm}$ is the momentum of the $D$ resonance in the $B$
rest frame and $\tilde{p}_{\nu}$ is the momentum of the neutrino in
the $\nu l$ rest frame, both of which are evaluated as
\begin{equation}
p_{D}^{\rm cm} = \frac{\lambda ^{1/2} (m_{B}^{2}, \, 
[M_{\rm inv}^{( \nu l )}]^{2}, \, m_{R}^{2}) }{2 m_{B}} , 
\end{equation}
\begin{equation}
\tilde{p}_{\nu} = \frac{\lambda ^{1/2} ([M_{\rm inv}^{( \nu l )}]^{2}, \, 
m_{\nu}^{2}, \, m_{l}^{2}) }{2 M_{\rm inv}^{( \nu l )}} ,
\label{eq:pnu_tilde}
\end{equation}
with the \Kaellen function $\lambda (x, \, y, \, z) = x^{2} + y^{2} +
z^{2} - 2 x y - 2 y z - 2 z x$ and the $D$ resonance mass $m_{R}$.
The tilde on characters indicates that they are evaluated in the $\nu
l$ rest frame unless explicitly mentioned.  The solid angles $\Omega
_{D}$ and $\tilde{\Omega} _{\nu}$ are for the $D$ resonance in the $B$
rest frame and for the neutrino in the $\nu l$ rest frame,
respectively, and $M_{\rm inv}^{(\nu l)}$ is the $\nu l$ invariant
mass.  After performing the angular integrals, we obtain the final
expression of the decay widths for the coalescence of the $D$
resonance:
\begin{equation}
\Gamma _{\rm coal} = 
\frac{| G_{\rm F} V_{b c} V_{\rm had} (D^{\ast}) |^{2}}{2 \pi ^{3} m_{B}^{2}}
\int d M_{\rm inv}^{( \nu l)} 
p_{D}^{\rm cm} \tilde{p}_{\nu} 
\overline{(E_{\nu} E_{l})_{B~\text{rest}}} , 
\label{eq:Gcoal}
\end{equation}
where the integral range of $M_{\rm inv}^{(\nu l)}$ is $[m_{l} +
m_{\nu} , \, m_{B} - m_{R} ]$.  In the equation, $\overline{(E_{\nu}
  E_{l})_{B~\text{rest}}}$ is the product of the $\nu$ and $l$
energies averaged over the neutrino solid angle and it is calculated in
the following way.  Before the angular integral, we have an
exact relation:
\begin{equation}
( E_{\nu} E_{l} )_{B~\text{rest}} 
= \frac{( p_{\nu} \cdot p_{B}) ( p_{l} \cdot p_{B})}{m_{B}^{2}} . 
\end{equation}
Since $( p_{\nu} \cdot p_{B})$ and $( p_{l} \cdot p_{B})$ are Lorentz
invariant, we may evaluate them in the $\nu l$ rest frame as
\begin{equation}
p_{\nu , l} \cdot p_{B} = \tilde{E}_{\nu , l} \tilde{E}_{B}
- \tilde{\bm{p}}_{\nu , l} \cdot \tilde{\bm{p}}_{B} ,
\quad 
\tilde{\bm{p}}_{\nu} = - \tilde{\bm{p}}_{l} .
\end{equation}
For simplicity we neglect lepton masses, so we have
\begin{equation}
\tilde{E}_{\nu} = \tilde{E}_{l} = 
\tilde{p}_{\nu} = \tilde{p}_{l} = \frac{M_{\rm inv}^{( \nu l)}}{2} 
\quad 
(\tilde{p}_{\nu , l} \equiv |\tilde{\bm{p}}_{\nu , l}|).
\end{equation}
On the other hand, the kinetic condition leads to an exact
form of $\tilde{E}_{B}$:
\begin{equation}
\tilde{E}_{B} = \frac{m_{B}^{2} + [ M_{\rm inv}^{( \nu l )}]^{2} 
- m_{R}^{2}}{2 M_{\rm inv}^{( \nu l )}} .
\end{equation}
In this way we have
\begin{equation}
p_{\nu , l} \cdot p_{B} = \frac{m_{B}^{2} + [ M_{\rm inv}^{( \nu l )}]^{2} - 
m_{R}^{2}}{4} 
- \tilde{\bm{p}}_{\nu , l} \cdot \tilde{\bm{p}}_{B} .
\end{equation}
Then $( E_{\nu} E_{l} )_{B~\text{rest}}$ becomes
\begin{align}
( E_{\nu} E_{l} )_{B~\text{rest}} & = 
\frac{\displaystyle \left ( 
\frac{m_{B}^{2} + [ M_{\rm inv}^{( \nu l )}]^{2} - 
m_{R}^{2}}{4}
- \tilde{\bm{p}}_{\nu} \cdot \tilde{\bm{p}}_{B}
\right )
\left ( \frac{m_{B}^{2} + [ M_{\rm inv}^{( \nu l )}]^{2} - 
m_{R}^{2}}{4}
+ \tilde{\bm{p}}_{\nu} \cdot \tilde{\bm{p}}_{B} \right )}
{m_{B}^{2}}
\notag \\
& = \left ( \frac{m_{B}^{2} + [M_{\rm inv}^{( \nu l)}]^{2}
- m_{R}^{2}}{4 m_{B}} \right ) ^{2} 
- \frac{( \tilde{\bm{p}}_{\nu} \cdot \tilde{\bm{p}}_{B} )^{2}}{m_{B}^{2}} .
\end{align}
Integrating the second term with the neutrino scattering angle in the
$\nu l$ rest frame, we have
\begin{equation}
- \frac{1}{2} \int _{-1}^{1} d \cos \tilde{\theta} _{\nu} 
\frac{( \tilde{\bm{p}}_{\nu} \cdot \tilde{\bm{p}}_{B} )^{2}}{m_{B}^{2}}
= - \frac{1}{3} \frac{( \tilde{p}_{\nu} \tilde{p}_{B} )^{2}}{m_{B}^{2}} ,
\end{equation}
where $\tilde{p}_{B} \equiv \sqrt{\tilde{E}_{B}^{2} - m_{B}^{2}}$.  As
a result, we obtain $\overline{(E_{\nu} E_{l})_{B~\text{rest}}}$ as
\begin{align}
\overline{( E_{\nu} E_{l} )_{B~\text{rest}}} 
& = \left ( \frac{m_{B}^{2} + [M_{\rm inv}^{( \nu l)}]^{2}
- m_{R}^{2}}{4 m_{B}} \right ) ^{2} 
- \frac{1}{3} \frac{( \tilde{p}_{\nu} \tilde{p}_{B} )^{2}}{m_{B}^{2}} .
\end{align}

\section{Rescattering}
\label{sec:rescattering}

Next, the production of two pseudoscalars with prompt production plus
rescattering through a $D$ resonance is calculated with the diagrams
shown in Fig.~6, and its hadronization amplitude $V_{\rm had}$ in the
isospin basis is given by
\begin{equation}
  V_{\text{had}} ( D K ) = C
  \left ( \sqrt{2} + \sqrt{2} G_{D K} T_{D K \to D K}
  - \frac{1}{\sqrt{3}} G_{D_{s} \eta} T_{D_{s} \eta \to D K} \right ) ,
\label{vhadk}
\end{equation}
\begin{equation}
  V_{\text{had}} ( D_{s} \eta ) = C
  \left ( - \frac{1}{\sqrt{3}} + \sqrt{2} G_{D K} T_{D K \to D_{s} \eta}
    - \frac{1}{\sqrt{3}} G_{D_{s} \eta} T_{D_{s} \eta \to D_{s} \eta} \right ) ,
\label{vhadeta}
\end{equation}
\begin{equation}
  V_{\text{had}} ( D \pi ) = C
  \left ( - \sqrt{\frac{3}{2}}
    - \sqrt{\frac{3}{2}} G_{D \pi} T_{D \pi \to D \pi}
  + G_{D_{s} \bar{K}} T_{D_{s} \bar{K} \to D \pi} \right ) ,
\label{vhadpi}
\end{equation}
\begin{equation}
  V_{\text{had}} ( D_{s} \bar{K} ) = C
  \left ( 1 - \sqrt{\frac{3}{2}} G_{D \pi} T_{D \pi \to D_{s} \bar{K}}
    + G_{D_{s} \bar{K}} T_{D_{s} \bar{K} \to D_{s} \bar{K}} \right ) .
\label{vhadsk}
\end{equation}
Again we see that the prefactor $C$ is the same in all the reactions.
In order to calculate decay widths in the particle basis, we have to
multiply by the appropriate Clebsch-Gordan coefficients.

Inserting Eq.~\eqref{vhadk} [or Eqs.~\eqref{vhadeta}, \eqref{vhadpi},
and \eqref{vhadsk}] into Eq.~\eqref{amp2} and the latter into
Eq.~\eqref{gamma4} we can derive the differential decay width $d\Gamma
_{i} / d M_{\rm inv}^{(i)}$, where $i$ represents the two pseudoscalar
states and $M_{\rm inv}^{(i)}$ is the invariant mass of the two
pseudoscalars, as
\begin{align}
\frac{d \Gamma _{i}}{d M_{\rm inv}^{(i)}}
= \frac{| G_{\rm F} V_{b c} V_{\rm had} (i) |^{2}}{8 \pi ^{5} m_{B}^{2}}
\int d M_{\rm inv}^{( \nu l)} 
P^{\rm cm} \tilde{p}_{\nu} \tilde{p}_{i}
\overline{(E_{\nu} E_{l})_{B~\text{rest}}} , 
\label{eq:dGamma_dM}
\end{align}
where $P^{\rm cm}$ is the momentum of the $\nu l$ system in the $B$
rest frame, $\tilde{p}_{\nu}$ is defined in Eq.~\eqref{eq:pnu_tilde},
and $\tilde{p}_{i}$ is the relative momentum of the two pseudoscalars
in their rest frame, both of which are evaluated as
\begin{equation}
P^{\rm cm} = \frac{\lambda ^{1/2} (m_{B}^{2}, \, 
[M_{\rm inv}^{( \nu l )}]^{2}, \, [M_{\rm inv}^{(i)}]^{2}) }{2 m_{B}} , 
\end{equation}
\begin{equation}
\tilde{p}_{i} = \frac{\lambda ^{1/2} ([M_{\rm inv}^{(i)}]^{2}, \, 
m_{i}^{2}, \, m_{i}^{\prime 2}) }{2 M_{\rm inv}^{(i)}} .
\end{equation}
Here we note that $\tilde{p}_{i}$ is a quantity in the rest frame of
the two pseudoscalars rather than the $\nu l$ system.

\section{The $\bm{D K}$-$\bm{D_{s} \eta}$ and $\bm{D \pi}$-$\bm{D_{s}
    \bar{K}}$ scattering amplitudes}
\label{sec:amplitudes}

In this section we will discuss in more detail the amplitudes which
appear in Eqs.~\eqref{vhadk}, \eqref{vhadeta}, \eqref{vhadpi}, and
\eqref{vhadsk}.  We formulate meson-meson scattering amplitudes for
the rescatterings to generate the $D_{s 0}^{\ast} (2317)$ and
$D_{0}^{\ast} (2400)$ resonances in the final state of the $B$ decay.
In Ref.~\cite{Gamermann:2006nm} it was found that the couplings to $D
K$ and $D_{s} \eta$ are dominant for $D_{s 0}^{\ast} (2317)$ and the
couplings to $D \pi$ and $D_{s} \bar{K}$ are dominant for
$D_{0}^{\ast} (2400)$.  Therefore, in the following we concentrate on
$D K$-$D_{s} \eta$ two-channel scattering in isospin $I=0$ and $D
\pi$-$D_{s} \bar{K}$ two-channel scattering in $I=1/2$, extracting
essential portions from Ref.~\cite{Gamermann:2006nm} and assuming
isospin symmetry.  Namely, we obtain these amplitudes by solving a
coupled-channel scattering equation in an algebraic form
\begin{equation}
T_{i j} (s) = V_{i j} (s) + \sum _{k} V_{i k} (s) G_{k} (s) T_{k j} (s) ,
\label{eq:BSEq}
\end{equation}
where $i$, $j$, and $k$ are channel indices, $s$ is the Mandelstam
variable of the scattering, $V$ is the interaction kernel, and $G$ is
the two-body loop function.

The interaction kernel $V$ corresponds to the tree-level transition
amplitudes  obtained from phenomenological Lagrangians developed in
Ref.~\cite{Gamermann:2006nm}.  Here we summarize the tree-level
amplitude in the isospin basis (for the two-body states in the isospin
basis, see Appendix~\ref{app:1}).  Namely, for the $D K$-$D_{s} \eta$
scattering in $I=0$ we have
\begin{equation}
V_{D K~D K}^{\rm phen.} (s, \, t, \, u)  = - \frac{1}{3 f_{\pi} f_{D}} \left [ \gamma ( t - u ) 
+ s - u + m_{D}^{2} + m_{K}^{2} \right ] ,
\end{equation}
\begin{equation}
 V_{D K~D_{s} \eta}^{\rm phen.} (s, \, t, \, u) = V_{\eta D_{s}~K D}^{\rm phen.} (s, \, t, \, u) 
 = - \frac{1}{6 \sqrt{3} f_{\pi} f_{D}} \big [ \gamma ( u - t ) - ( 3 + \gamma ) ( s - u ) 
 - m_{D}^{2} - 3 m_{K}^{2} + 2 m_{\pi}^{2} \big ] ,
\end{equation}
\begin{equation}
V_{D_{s} \eta~D_{s} \eta}^{\rm phen.} (s, \, t, \, u)  = - \frac{1}{9 f_{\pi} f_{D}} \left [ \gamma ( - s + 2 t - u ) 
+ 2 m_{D}^{2} + 6 m_{K}^{2} - 4 m_{\pi}^{2} \right ] ,
\end{equation}
and for the $D \pi$-$D_{s} \bar{K}$ scattering in $I=1/2$ we have
\begin{equation}
V_{D \pi~D \pi}^{\rm phen.} (s, \, t, \, u) \notag  = - \frac{1}{12 f_{\pi} f_{D}} \big [ 2 \gamma ( t - u ) 
+ ( \gamma + 4 ) ( s - u )  + 2 m_{D}^{2} + 2 m_{\pi}^{2} \big ] ,
\end{equation}
\begin{equation}
V_{D \pi~D_{s} \bar{K}}^{\rm phen.} (s, \, t, \, u) = V_{\bar{K} D_{s}~\pi D}^{\rm phen.} (s, \, t, \, u) 
= \frac{1}{2 \sqrt{6} f_{\pi} f_{D}} \left [ \gamma ( t - u ) + s - u  + m_{D}^{2} + m_{K}^{2} \right ] .
\end{equation}
\begin{equation}
V_{D_{s} \bar{K}~D_{s} \bar{K}}^{\rm phen.} (s, \, t, \, u) = - \frac{1}{6 f_{\pi} f_{D}} \left [ \gamma ( t - u ) + s - u
+ m_{D}^{2} + 2 m_{K}^{2} - m_{\pi}^{2} \right ] .
\end{equation}
where $t$ and $u$ are Mandelstam variables. In these equations,
$f_{\pi}$ and $f_{D}$ represent the pion and $D$ meson decay
constants, respectively, and $m_{\pi}$, $m_{K}$, and $m_{D}$ are the
masses of pion, kaon, and $D$ mesons, respectively.  In addition, in
order to treat effectively interactions of heavy mesons, we have
introduced a parameter $\gamma \equiv ( m_{\rm L} / m_{\rm H} )^{2}$
as the squared ratio of the masses of the light to heavy vector mesons
(respectively $m_{\rm L}$ and $m_{\rm H}$), which are exchanged
between two pseudoscalar mesons.  Then we perform the on-shell
factorization and the $s$-wave projection to give the interaction
kernel $V$ in Eq.~\eqref{eq:BSEq}:
\begin{equation}
V (s) = \frac{1}{2} \int _{-1}^{1} d \cos \theta \, V^{\rm phen.} 
(s, \, t ( s, \, \cos \theta ), \, u ( s, \, \cos \theta )) ,
\end{equation}
where $\theta$ is the scattering angle in the center-of-mass frame.

For the loop function $G$, on the other hand, we use the expression in
Eq.~\eqref{eq:Gloop}.  In this study we employ the dimensional
regularization, so we can express the loop function as
\begin{equation}
G_{k} (s) = \frac{1}{16 \pi ^{2}} \left [ a_{k} ( \mu _{\rm reg} ) 
+ \ln \frac{m_{k}^{2}}{\mu _{\rm reg}^{2}} 
+ \frac{s + m_{k}^{\prime 2} - m_{k}^{2}}{2 s} 
\ln \frac{m_{k}^{\prime 2}}{m_{k}^{2}} 
- \frac{2 \lambda ^{1/2} (s, \, m_{k}^{2}, \, m_{k}^{\prime 2})}{s} 
\text{artanh} 
\left ( \frac{\lambda ^{1/2} (s, \, m_{k}^{2}, \, m_{k}^{\prime 2})}
{m_{k}^{2} + m_{k}^{\prime 2} - s} \right ) 
\right ] ,
\label{eq:Gdim}
\end{equation}
with the regularization scale $\mu _{\rm reg}$ and the subtraction
constant $a_{k}$, which becomes a model parameter.
In this approach, $D$ resonances can appear as poles of the
scattering amplitude $T_{i j} (s)$ with the residue $g_{i} g_{j}$:
\begin{equation}
T_{i j} ( s ) = \frac{g_{i} g_{j}}{s - s_{\rm pole}}
+ ( \text{regular at }s = s_{\rm pole} ) . 
\label{eq:amp_pole}
\end{equation}
The pole is described by its position $s_{\rm pole}$ and the constant
$g_{i}$, which can be interpreted as the coupling constant of the $D$
resonance to the $i$ channel. In this study only the subtraction constant in 
each channel is the model parameter.  Actually, the meson masses are fixed as 
$m_{\pi} = 138.04 \mev$, $m_{K} = 495.67 \mev$, $m_{\eta} = 547.85 \mev$, $m_{D}
= 1867.23 \mev$, and $m_{D_{s}} = 1968.30 \mev$, and we take
\begin{equation}
f_{\pi} = 93 \mev , 
\quad 
f_{D} = 165 \mev , 
\end{equation}
\begin{equation}
m_{\rm L} = 800 \mev , 
\quad 
m_{\rm H} = 2050 \mev ,
\end{equation}
for the pion and $D$ decay constants and masses of the light and heavy
vector mesons, respectively.  On the other hand, the subtraction
constant, as a model parameter, is determined so as to generate a pole
of $D_{s 0}^{\ast} (2317)$ at the right place, i.e., to reproduce the
mass reported by the Particle Data Group from the square root of the
pole position, $\sqrt{s_{\rm pole}}$.  In this study we assume that
all the subtraction constants take the same value for simplicity, and
employ $a_{D K} = a_{D_{s} \eta} = a_{D \pi} = a_{D_{s} \bar{K}} = -
1.27$ at $\mu _{\rm reg} = 1500 \mev$.  Indeed, with these values of
the subtraction constant we obtain the pole positions listed in
Table~\ref{tab:Ds}.  The values of the coupling constants $g_{i}$ are
also given in Table~\ref{tab:Ds}.

\begin{table}
  \caption{Pole position $\sqrt{s_{\rm pole}}$, coupling constant
    $g_{i}$, compositeness $X_{i}$, and elementariness $Z$ for the $D$
    resonances in the isospin basis.  }
  \label{tab:Ds}
  % \begin{ruledtabular}
  \begin{tabular*}{8.6cm}{@{\extracolsep{\fill}}lcc|lc}
    \hline \hline
    \multicolumn{2}{c}{$D_{s 0}^{\ast} (2317)$} &  &
    \multicolumn{2}{c}{$D_{0}^{\ast} (2400)$} \\
    \hline 
    $\sqrt{s_{\rm pole}}$ & $2317 \mev$ & &
    $\sqrt{s_{\rm pole}}$ & $2128 - 160 i \mev$ \\
    $g_{D K}$ & $10.58 \gev$ &  &
    $g_{D \pi}$ & $\phph 9.00 - 6.18 i \gev$ \\
    $g_{D_{s} \eta}$ & $- 6.11 \gev ~$ & &
    $g_{D_{s} \bar{K}}$ & $- 7.68 + 4.35 i \gev$ \\
    $X_{D K}$ & $0.69$ & &
    $X_{D \pi}$ & $0.34 + 0.41 i$ \\
    $X_{D_{s} \eta}$ & $0.09$ & &
    $X_{D_{s} \bar{K}}$ & $0.03 - 0.12 i$ \\
    $Z$ & $0.22$ & &
    $Z$ & $0.63 - 0.28 i$ \\
    \hline \hline
  \end{tabular*}
  % \end{ruledtabular}
\end{table}

As one can see from Table~\ref{tab:Ds}, the $D_{s 0}^{\ast} (2317)$
state has zero decay width with $\text{Im} \sqrt{s_{\rm pole}} = 0$,
since we do not include the $D_{s} \pi ^{0}$ decay channel, for which
the isospin symmetry breaking is necessary.  As a result, the coupling
constants also become real and have positive values.  The $D K$ coupling
constant is about two times larger than that of the $D_{s} \eta$
coupling constant, and their values are in agreement with the results
obtained in Ref.~\cite{Gamermann:2006nm}.  On the other hand, the
$D_{0}^{\ast} (2400)$ state has a decay width ($2 \, \text{Im}
\sqrt{s_{\rm pole}} \approx 320 \mev$) to the $D \pi$ decay channel,
and the coupling constant is complex.  The magnitude of the $D \pi$
coupling constant is larger than that of the $D_{s} \bar{K}$ coupling
constant, and they are very close to the values in
Ref.~\cite{Gamermann:2006nm}.  In the following we will use the
coupling constants of the $D$ resonances in Table~\ref{tab:Ds} for the
coalescence of $D$ resonances in the semileptonic $B$ decays [see
Eq.~\eqref{eq:Gcoal}] and use the scattering amplitude for the
meson-meson invariant mass distributions of the $D$ resonances [see
Eq.~\eqref{eq:dGamma_dM}].

Let us further discuss the structure of the $D$ resonances in this
model from the point of view of compositeness, which is defined as the
contribution from the two-body part to the normalization of the total
wave function and measures the fraction of the two-body
state~\cite{Hyodo:2011qc, Aceti:2012dd, Xiao:2012vv, Hyodo:2013nka,
  Sekihara:2014kya}.  Actually, the coupling constant $g_{i}$ is found
to be the coefficient of the two-body wave function in
Refs.~\cite{Gamermann:2009uq, YamagataSekihara:2010pj}, and the
expression of the compositeness in the present model
is%~\cite{Hyodo:2011qc, Hyodo:2013nka, Sekihara:2014kya}
\begin{equation}
X_{i} = 
- g_{i}^{2} \left [ \frac{d G_{i}}{d s} \right ] _{s = s_{\rm pole}} .
\end{equation}
On the other hand, the elementariness $Z$, which measures the fraction
of missing channels, are expressed as
\begin{equation}
Z = - \sum _{i, j} g_{j} g_{i} \left [ G_{i} \frac{d V_{i j}}{d s} 
G_{j} \right ] _{s = s_{\rm pole}} .
\end{equation}
We note that in general both the compositeness $X_{i}$ and
elementariness $Z$ become complex values for a resonance state and
hence one cannot interpret the compositeness (elementariness) as the
probability to observe a two-body (missing-channel) component inside
the resonance.  However, a striking property is that the sum of them
coincides with the normalization of the total wave function for the
resonance and is exactly unity:
\begin{equation}
\sum _{i} X_{i} + Z = 1 ,
\end{equation}
which is guaranteed by a generalized Ward identity proved in
Ref.~\cite{Sekihara:2010uz}.  Therefore one can deduce the structure
by comparing the value of the compositeness with unity, on the basis
of the similarity to the stable bound state case.  The values of the
compositeness and elementariness of the $D$ resonances in this
approach are also listed in Table~\ref{tab:Ds}.  The result indicates
that the $D_{s 0}^{\ast} (2317)$ resonance, which is obtained as a
bound state in the present model, is indeed dominated by the $D K$
component.  This has been corroborated in the recent analysis of QCD
lattice results of~\cite{sasa}. In contrast, we may interpret that the
$D_{0}^{\ast} (2400)$ resonance is constructed with missing channels,
although the imaginary part for each component is not negligible.

\section{Numerical results}
\label{sec:results}

Now we show our numerical results of the semileptonic $B$ decay
widths.  As we have seen, we fix the hadronization process of the two
mesons in Sec.~\ref{sec:hadronization} and we employ an effective
model in Sec.~\ref{sec:amplitudes} so as to determine the strength of
the couplings of the $D$ resonances to the meson-meson channels.  In
this way, we can calculate the ratio of the decay widths in the
coalescence treatment as well as in the rescattering.

%\subsection{Branching fraction}

\begin{table}
  \caption{Ratios of decay widths and branching fractions 
    of semileptonic $B$ decays. }
  \label{tab:Gamma3}  
  % \begin{tabular*}{\textwidth}{@{\extracolsep{\fill}}lcc}
  \begin{tabular*}{8.6cm}{@{\extracolsep{\fill}}lc}
    \hline \hline
%    & Present study \\
%    \hline
    $R$ & $0.45$ \\
    $\Gamma_{B^{-} \to D_{0}^{\ast}(2400)^{0} \bar{\nu}_{l} l^{-}} 
    / \Gamma_{\bar{B}^{0} \to D_{0}^{\ast}(2400)^{+} \bar{\nu}_{l} l^{-}}$ & 
    $1.00$ \\
%
%    $\mathcal{B}[\bar{B}_{s}^{0} \to D_{s0}^{\ast}(2317)^{+} \bar{\nu}_{l} l^{-}]$ &    $1.3 \times 10^{-3}$ \\
    $\mathcal{B}[\bar{B}^{0} \to D_{0}^{\ast} (2400)^{+} \bar{\nu} _{l} l^{-}]$     &    $3.0 \times 10^{-3}$ (input) \\
    $\mathcal{B}[\bar{B}^{-} \to D_{0}^{\ast} (2400)^{0} \bar{\nu} _{l} l^{-}$] &    $3.2 \times 10^{-3}$ \\
    \hline \hline
  \end{tabular*}
\end{table}

First we consider the coalescence case.  The numerical results are
summarized in Table~\ref{tab:Gamma3}.  The most interesting quantity
is the ratio $R = \Gamma_{\bar{B}_{s}^{0} \to D_{s0}^{\ast} (2317)^{+}
  \bar{\nu}_{l} l^{-}} / \Gamma_{\bar{B}^{0} \to D_{0}^{\ast}
  (2400)^{+} \bar{\nu}_{l} l^{-}}$ in the coalescence treatment, which
removes the unknown factor $C$ in the hadronization process.  The
decay width in the coalescence is expressed in Eq.~\eqref{eq:Gcoal}.
The coupling constants of the two mesons to the $D$ resonances are
determined in Sec.~\ref{sec:amplitudes} and listed in
Table~\ref{tab:Ds}.  We emphasize that we have no fitting parameters
for the ratio $R$ in this scheme.  As a result, we obtain the ratio of
the decay widths as $R = 0.45$.  On the other hand, we find that the
ratio $\Gamma_{B^{-} \to D_{0}^{\ast}(2400)^{0} \bar{\nu}_{l} l^{-}} /
\Gamma_{\bar{B}^{0} \to D_{0}^{\ast}(2400)^{+} \bar{\nu}_{l} l^{-}}$
is $1.00$, which can be expected from the same strength of the decay
amplitude to the charged and neutral $D_{0}^{\ast} (2400)$ due to the
isospin symmetry, as discussed after Eq.~\eqref{eq:phiphi_41}.

Then, we can fix the absolute value of the common prefactor $C$ by
using experimental data of the decay width.  Actually, the branching
fraction of the semileptonic decay $\bar{B}^{0} \to D_{0}^{\ast}
(2400)^{+} \bar{\nu} _{l} l^{-}$ to the total decay is reported as
$(3.0 \pm 1.2) \times 10^{-3}$ by the Particle Data
Group~\cite{Agashe:2014kda}.  By using this mean value we find $C =
7.28$, and the fractions of decays $\bar{B}_{s}^{0} \to
D_{s0}^{\ast}(2317)^{+} \bar{\nu}_{l} l^{-}$ and $B^{-} \to
D_{0}^{\ast} (2400)^{0} \bar{\nu} _{l} l^{-}$ to the total decay
widths are obtained as $1.3 \times 10^{-3}$ and $3.2 \times 10^{-3}$,
respectively.  The values of these fractions are similar to each other.
The difference of the fractions of $\bar{B}^{0} \to D_{0}^{\ast}
(2400)^{+} \bar{\nu} _{l} l^{-}$ and $B^{-} \to D_{0}^{\ast}
(2400)^{0} \bar{\nu} _{l} l^{-}$ comes from the fact that the total
decay widths of $\bar{B}^{0}$ and $B^{-}$ are different.
\begin{table}
  \caption{Branching fraction of the process $\bar{B}_{s}^{0} \to D_{s0}^{\ast}(2317)^{+} \bar{\nu}_{l} l^{-}$ in percentage}
  \label{tab:bran2}  
  % \begin{tabular*}{\textwidth}{@{\extracolsep{\fill}}lcc}
  \begin{tabular*}{8.6cm}{@{\extracolsep{\fill}}lc}
    \hline \hline
%    & Present study \\
%    \hline
     Approach  &   $\mathcal{B}[\bar{B}_{s}^{0} \to D_{s0}^{\ast}(2317)^{+} \bar{\nu}_{l} l^{-}]$ \\
     \hline
     This work &     $0.13$ \\
     QCDSR + HQET \cite{huang} &  $0.09 - 0.20$ \\
     QCDSR (SVZ)  \cite{aliev} &  $0.10$  \\
     LCSR         \cite{li}    &  $0.23 \pm 0.11$\\
     CQM          \cite{zhao}  &  $0.49 - 0.57$ \\ 
     CQM          \cite{seg}   &  $0.44$ \\
     CQM          \cite{alb}   &  $0.39$ \\  
     \hline \hline
  \end{tabular*}
\end{table}

In Table \ref{tab:bran2} we compare our predictions for
$\mathcal{B}[\bar{B}_{s}^{0} \to D_{s0}^{\ast}(2317)^{+} \bar{\nu}_{l}
l^{-}]$ with the results obtained with other approaches. Although not
explicitly mentioned by the authors, from the reading of the works we
can see that we should attach an uncertainty of at least $10$ \% to
the numbers without theoretical error bars.  The discrepancy between
the calculated branching fractions can be of a factor five, showing
that there is a large room for improvement on the theoretical side.
Our approach is the only one where the $D_{s0}^{\ast}(2317)^{+}$ is
treated as a mesonic molecule. Looking at Table \ref{tab:bran2} we can
divide the results in two groups: the first four numbers, which are
``small'' and the last three, which are ``large''. In the second
group, the constituent quark models (CQM) yield larger branching
fractions.  Understanding the origin of the discrepancies requires a
very careful comparative analysis of all the ingredients of the
different approaches and it is beyond the scope of the present work.
However it is tempting, as a first speculation, to attribute these
differences to the differences in spatial configurations, which are
inherent to each approach. The parent $B_s$ meson is a compact state,
with a typical radius of the order of the lowest charmonium radius,
i.e.  $\langle r \rangle \simeq 0.4$ fm. In constituent quark models
all the $D_s$ mesons, being relatively heavy $c \bar{q}$ states,
should also be compact and hence the overlap between the initial and
final state spatial wave functions is large. In the molecular picture
of the $D_{s0}^{\ast}(2317)^{+}$, a bound state of two mesons is
expected to have a large radius, of the order of a few fm, and
therefore in this case the overlap between initial and final of wave
functions is small, reducing the corresponding branching fraction. In
the QCD sum rules formalism, there is no explicit mention to the
spatial configuration of the interpolating currents and it is
difficult to say anything.

%\subsection{Invariant mass distributions}

Next we consider the rescattering process for the final-state two
mesons formulated in Sec.~\ref{sec:rescattering}.  We use the common
prefactor $C = 7.28$ fixed from the experimental value of the width of
the semileptonic decay $\bar{B}^{0} \to D_{0}^{\ast} (2400)^{+}
\bar{\nu} _{l} l^{-}$.  The meson-meson scattering amplitude is
obtained in Sec.~\ref{sec:amplitudes}, and we further introduce the
$D_{s} \pi ^{0}$ channel as the isospin-breaking decay mode of
$D_{s0}^{\ast} (2317)$.  Namely, we calculate the scattering amplitude
involving the $D_{s} \pi ^{0}$ channel as
\begin{equation}
T_{i \to D_{s} \pi ^{0}} = \frac{g_{i} g_{D_{s} \pi ^{0}}}
{s - [ M_{D_{s0}^{\ast}} - i \Gamma _{D_{s0}^{\ast}} / 2]^{2}} ,
\end{equation}
for $i = D K$ and $D_{s} \eta$.  We take the $D_{s0}^{\ast} (2317)$
mass as $M_{D_{s0}^{\ast}} = 2317 \mev$, while we assume its decay
width as $\Gamma _{D_{s0}^{\ast}} = 3.8 \mev$, which is the upper
limit from experiments~\cite{Agashe:2014kda}.  The $D_{s0}^{\ast}
(2317)$-$i$ coupling constant $g_{i}$ ($i = D K$, $D_{s} \eta$) is
taken from Table~\ref{tab:Ds}, and the $D_{s0}^{\ast} (2317)$-$D_{s}
\pi ^{0}$ coupling constant $g_{D_{s} \pi ^{0}}$ is calculated from
the $D_{s0}^{\ast} (2317)$ decay width as
\begin{equation}
g_{D_{s} \pi ^{0}} = \sqrt{\frac{8 \pi M_{D_{s0}^{\ast}}^{2} 
\Gamma _{D_{s0}^{\ast}}}{p_{\pi}}} ,
\end{equation}
with the pion center-of-mass momentum $p_{\pi}$, and we obtain
$g_{D_{s} \pi ^{0}} = 1.32 \gev$.

\begin{figure}[!t]
  \centering
%  \Psfig{8.6cm}{dGdM.eps}
  \Psfig{8.6cm}{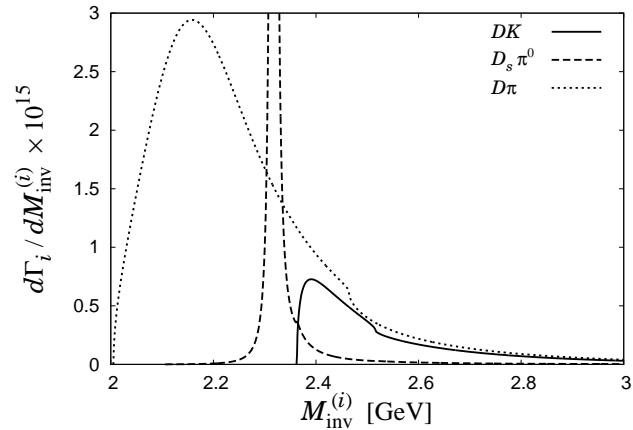}
  \caption{Differential decay width $d \Gamma _{i} / d M_{\rm
      inv}^{(i)}$ for the two pseudoscalars channel $i$ in the isospin
    basis.  Here we consider the semileptonic decays $\bar{B}_{s}^{0}
    \to (D K)^{+} \bar{\nu}_{l} l^{-}$, $(D_{s} \pi ^{0})^{+}
    \bar{\nu}_{l} l^{-}$ and $\bar{B}^{0} \to (D \pi )^{+} \bar{\nu}
    _{l} l^{-}$.  The $D K$ and $D_{s} \pi ^{0}$ channels couple to
    the $D_{s0}^{\ast}(2317)^{+}$ resonance, and $D \pi$ to the
    $D_{0}^{\ast} (2400)$ resonance.  The peak height for the $D_{s}
    \pi ^{0}$ channel is $d \Gamma _{D_{s} \pi ^{0}} / d M_{\rm
      inv}^{(D_{s} \pi ^{0})} \sim 10^{-13}$.  }
  \label{fig7}
\end{figure}

The results of the differential decay width $d\Gamma _{i} / d M_{\rm
  inv}^{(i)}$~\eqref{eq:dGamma_dM}, where $i$ represents the two
pseudoscalar states, are shown in Fig.~\ref{fig7}.  The figure
is plotted in the isospin basis.  Therefore, when translating into the
particle basis we use the relation according to the weight of states
given in Appendix~\ref{app:1}:
\begin{equation}
[ D^{0} K^{+} ] = [ D^{+} K^{0} ] = \frac{1}{2} [ D K ] ,
\end{equation}
\begin{equation}
[ D_{s}^{+} \pi ^{0} ] = [ D_{s} \pi ^{0} ] ,
\end{equation}
\begin{equation}
[ D^{0} \pi ^{+} ] = 2 [ D^{+} \pi ^{0} ] = \frac{2}{3} [ D \pi ] ,
\end{equation}
where $[A B]$ is the partial decay width to the $A B$ channel.  An
interesting point is that the $D K$ mode shows a rapid increase from
its threshold $\approx 2360 \mev$ due to the existence of the bound
state, i.e., the $D_{s0}^{\ast} (2317)$ resonance.  In experiments,
such a rapid increase from the $D K$ threshold would support the
interpretation of the $D_{s0}^{\ast} (2317)$ resonance as a $D K$
bound state.  The strength of the $D K$ contribution in the $M_{\rm
  inv}^{(i)} \gtrsim 2.4 \gev$ region is similar to that of $D \pi$,
which corresponds to the ``tail'' for the $D_{0}^{\ast} (2400)$
resonance.  In fact, the position of the $D_{0}^{\ast} (2400)$ peak in
Fig.~\ref{fig7} might be shifted to higher invariant masses, since
we have underestimated the mass of the $D_{0}^{\ast} (2400)$ resonance
in our model compared to the experimental values $2318 \mev$ and $2403
\mev$ for neutral and charged $D_{0}^{\ast} (2400)$,
respectively~\cite{Agashe:2014kda} (note that the experimental
uncertainties in the position and width of this resonance are large).
On the other hand, the $D_{s} \pi ^{0}$ peak coming from the
$D_{s0}^{\ast} (2317)$ resonance is very sharp due to its narrow
width.  The $D_{s} \pi ^{0}$ peak height is about 30 times larger than
the $D \pi$ one coming from $D_{0}^{\ast} (2400)$, but when
integrating the bump structure of the differential decay widths we
obtain a ratio of semileptonic $B$ decays into $D_{s0}^{\ast} (2317)$
to $D_{0}^{\ast} (2400)$ close to $0.45$, as obtained in the
coalescence treatment above.

The spectra shown in Fig.~\ref{fig7} are our predictions and they may be measured at the LHCb.
They were obtained in the framework of the chiral unitarity approach in coupled channels and their 
experimental observation would give support to the $D_{s0}^{\ast} (2317)$ and $D_{0}^{\ast} (2400)$ 
as dynamically generated resonances, which is inherent to this approach.

\section{Conclusion}
\label{sec:conclusion}

We have extended the formalism developed in \cite{liang} and applied
it to semileptonic $B$ and $B_s$ decays into resonances, which are
interpreted as dynamically generated resonances. As in \cite{liang},
we start studying the weak process at the quark level and, as a
``final state interaction'', the outgoing quark-antiquark pair couples
to meson pairs, which rescatter and form resonances, which then decay
in well defined channels. This process, discussed in
Sec.~\ref{sec:hadronization}, is a very economic hadronization
mechanism with a single parameter, $C$.  After fixing it with the help
of experimental information on the $\bar{B}^{0} \to
D_{0}^{\ast}(2400)^{+} \bar{\nu} _{l} l^{-}$ decay, we make
predictions for the semileptonic decay width of the $D_{s0}^{\ast}
(2317)$, shown in Table \ref{tab:bran2} and also for the invariant
mass spectra shown in Fig.~\ref{fig7}.

We have added a new information related to the nature of the
$D_{s0}^{\ast} (2317)$ as an object with a dominant $D K$ molecular
component, which is the $D K$ mass distribution in the
$\bar{B}_{s}^{0} \to (D K)^{+} \bar{\nu}_{l} l^{-}$ decay.  The
simultaneous measurement of the decay rate into the $D_{s0}^{\ast}
(2317)$ resonance and the related $D K$ mass distribution are hence
strongly encouraged to gain further knowledge on the nature of this
resonance. The experimental confirmation of our predictions would give
additional support to the $D_{s0}^{\ast} (2317)$ and $D_{0}^{\ast}
(2400)$ resonances as dynamically generated resonances from the
meson-meson interaction.

\begin{acknowledgments}

  We thank A.~Hosaka, A.~M.~Torres and K.~Khemchandani for useful
  discussions.  We acknowledge the support by FAPESP, CNPq and by Open
  Partnership Joint Projects of JSPS Bilateral Joint Research
  Projects.  This work is partly supported by the Spanish Ministerio
  de Economia y Competitividad and European FEDER funds under the
  contract number FIS2011-28853-C02-01 and FIS2011-28853-C02-02, and
  the Generalitat Valenciana in the program Prometeo II-2014/068.  We
  acknowledge the support of the European Community-Research
  Infrastructure Integrating Activity Study of Strongly Interacting
  Matter (acronym HadronPhysics3, Grant Agreement n. 283286) under the
  Seventh Framework Program of the EU. F.S.N. and M.N.  are deeply
  grateful to the members of the IFIC/Universitat de Valencia for the
  hospitality and support extended to them during a visit in November,
  2014.

\end{acknowledgments}

\appendix 

\section{Conventions}
\label{app:1}

In this Appendix we show conventions used in this study.  Throughout
this article we employ the metric in four-dimensional Minkowski space
defined as $g^{\mu \nu} = g_{\mu \nu} = \text{diag}(1, \, -1, \, -1,
\, -1)$ and the Einstein summation convention is used unless
explicitly mentioned.

We introduce the Dirac spinors $u ( \bm{p}, \, s)$ and $v ( \bm{p}, \,
s)$, where $\bm{p}$ is three-momentum of the field and $s$ represents
its spin, as the positive and negative energy solutions of the Dirac
equation, respectively:
\begin{equation}
( \Slash{p} - m ) u(\bm{p} , \, s) = 0 , 
\quad 
( \Slash{p} + m ) v(\bm{p} , \, s) = 0 . 
\end{equation}
Here $m$ is the mass of the field, $\Slash{p} \equiv \gamma ^{\mu}
p_{\mu}$ with $\gamma ^{\mu}$ being the Dirac gamma matrices, and
$p^{\mu} \equiv \left ( \sqrt{\bm{p}^{2} + m^{2}}, \, \bm{p} \right )$
is the on-shell four-momentum of the solution.  In this study the
Dirac spinors are normalized as follows:
\begin{equation}
\overline{u}(\bm{p}, \, s) u(\bm{p}, \, s^{\prime}) 
= \delta _{s s^{\prime}} , 
\quad 
\overline{v}(\bm{p}, \, s) v(\bm{p}, \, s^{\prime}) 
= - \delta _{s s^{\prime}} , 
\end{equation}
with $\overline{u} \equiv u^{\dagger} \gamma ^{0}$ and $\overline{v}
\equiv v^{\dagger} \gamma ^{0}$, and hence we have
\begin{equation}
\begin{split}
& \sum _{s} 
u(\bm{p}, \, s) \overline{u}(\bm{p}, \, s) 
= \frac{\Slash{p} + m}{2 m} , 
\\
& \sum _{s} 
v(\bm{p}, \, s) \overline{v}(\bm{p}, \, s) 
= \frac{\Slash{p} - m}{2 m} . 
\end{split}
\end{equation}
The trace identities used in this study is summarized as follows:
\begin{equation}
\trace \left [ \gamma ^{\mu} \gamma ^{\nu} 
\gamma ^{\rho} \gamma ^{\sigma} \right ]
= 4 ( g^{\mu \nu} g^{\rho \sigma} - g^{\mu \rho} g^{\nu \sigma} 
+ g^{\mu \sigma} g^{\nu \rho} ) ,
\end{equation}
\begin{equation}
\trace \left [ \gamma _{5} \gamma ^{\mu} \gamma ^{\nu} 
\gamma ^{\rho} \gamma ^{\sigma} \right ]
= - 4 i \epsilon ^{\mu \nu \rho \sigma} ,
\end{equation}
\begin{equation}
\trace \left [ \gamma  ^{\mu} \gamma ^{\nu} 
\gamma ^{\rho} \right ] 
= \trace \left [ \gamma _{5} \gamma  ^{\mu} \gamma ^{\nu} 
\gamma ^{\rho} \right ] 
= 0 , 
\end{equation}
where $\gamma _{5} \equiv i \gamma ^{0} \gamma ^{1} \gamma ^{2} \gamma
^{3}$ and $\epsilon ^{\mu \nu \rho \sigma}$ is the Levi-Civita symbol
with the normalization $\epsilon ^{0 1 2 3} = 1$.

The phase convention for mesons in terms of the isospin states $| I ,
\, I_{3} \rangle$ used in this study is given by
\begin{equation}
\begin{split}
& | \pi ^{+} \rangle = - | 1, \, 1 \rangle , 
\quad 
| K^{-} \rangle 
= - | 1/2, \, - 1/2 \rangle , 
\\
& | D^{0} \rangle 
= - | 1/2 , \, - 1/2 \rangle ,
\end{split}
\end{equation}
while other meson states in this study are represented without phase
factors.  As a result, we can translate the two-body states used in
this study into the isospin basis, which we specify as $(I, \,
I_{3})$, as
\begin{equation}
| D K (0, \, 0) \rangle = 
\frac{1}{\sqrt{2}} | D^{0} K^{+} \rangle 
+ \frac{1}{\sqrt{2}} | D^{+} K^{0} \rangle , 
\end{equation}
\begin{equation}
| D_{s} \eta (0, \, 0) \rangle = 
| D_{s}^{+} \eta \rangle , 
\end{equation}
\begin{equation}
| D \pi (1/2, \, 1/2) \rangle = 
- \sqrt{\frac{2}{3}} | D^{0} \pi ^{+} \rangle 
+ \frac{1}{\sqrt{3}} | D^{+} \pi ^{0} \rangle , 
\end{equation}
\begin{equation}
| D_{s} \bar{K} (1/2, \, 1/2) \rangle = 
| D_{s}^{+} \bar{K}^{0} \rangle , 
\end{equation}
\begin{equation}
| D \pi (1/2, \, -1/2) \rangle = 
+ \frac{1}{\sqrt{3}} | D^{0} \pi ^{0} \rangle 
+ \sqrt{\frac{2}{3}} | D^{+} \pi ^{-} \rangle , 
\end{equation}
\begin{equation}
| D_{s} \bar{K} (1/2, \, -1/2) \rangle = 
- | D_{s}^{+} K^{-} \rangle . 
\end{equation}

At last we summarize the Feynman rules for the weak interaction used
in this study.  We express the $W \nu l$ coupling as
\begin{equation}
- i V_{W \nu l}^{\mu} = i \frac{g_{\rm W}}{\sqrt{2}} \gamma ^{\mu}
\frac{1 - \gamma _{5}}{2} ,
\end{equation}
with $g_{\rm W}$ being the coupling constant of the weak interaction
and the $W b c$ coupling as
\begin{equation}
- i V_{W \nu l}^{\mu} = i \frac{g_{\rm W} V_{b c}}{\sqrt{2}} \gamma ^{\mu}
\frac{1 - \gamma _{5}}{2} ,
\end{equation}
where $V_{b c}$ is the Cabibbo-Kobayashi-Maskawa matrix elements,
whose absolute value is $| V_{b c} | \approx 0.041$.  The $W$ boson
propagator with four-momentum $p^{\mu}$ is written as
\begin{equation}
i P_{W}^{\mu \nu} ( p ) =  \frac{- i g^{\mu \nu}}{p^{2} - M_{W}^{2} + i 0} ,
\end{equation}
with the mass of the $W$ boson $M_{W}$.  The coupling constant $g_{\rm
  W}$ and the mass of the $W$ boson $M_{W}$ are related to the Fermi
coupling constant $G_{\rm F}$ as
\begin{equation}
G_{\rm F} = \frac{g_{\rm W}^{2}}{4 \sqrt{2} M_{W}^{2}} 
\approx 1.166 \times 10^{-5} \gev ^{-2} .
\end{equation}

\end{document}